\documentclass{aa}  

\usepackage{graphicx}
\usepackage{natbib}
\bibpunct{(}{)}{;}{a}{}{,}
\usepackage{txfonts}
\usepackage[
  breaklinks = true,
  colorlinks = true,
  urlcolor   = blue,
  citecolor  = blue,
  linkcolor  = blue,
]{hyperref}
\usepackage{xcolor}

\newcommand{\Ha}{\textnormal{H}\alpha}

\newcommand{\CaII}{\textnormal{Ca}~\textsc{ii}}
\newcommand{\kms}{\textnormal{km s}^{-1}}

\defcitealias{2025A&A...697A.180S}{Paper~I}
\defcitealias{2017Sci...356.1269M}{MS17}
\defcitealias{2020ApJ...889...95M}{MS20}

\begin{document} 

    \title{Shock-induced magnetic reconnection driving\\
    Ellerman bomb emission and a spicule}

    \titlerunning{Shock-induced magnetic reconnection driving Ellerman bomb emission and a spicule}

    \authorrunning{M.O. Sand et al.}
    
    \author{
        Mats Ola Sand
        \inst{1, 2},
        Quentin Noraz
        \inst{1, 2},
        Guillaume Aulanier
        \inst{3, 1, 2},
        Juan Martínez-Sykora
        \inst{4, 5, 1, 2}
        \\
        Mats Carlsson
        \inst{1, 2},
        Luc Rouppe van der Voort
        \inst{1, 2}
        }
   \institute{
        Institute of Theoretical Astrophysics, University of Oslo, P.O. Box 1029 Blindern, N-0315 Oslo, Norway
        \email{m.o.sand@astro.uio.no}
        \and
        Rosseland Centre for Solar Physics, University of Oslo, P.O. Box 1029 Blindern, N-0315 Oslo, Norway
        \and
        Sorbonne Université, Observatoire de Paris – PSL, École Polytechnique, Institut Polytechnique de Paris, CNRS, Laboratoire de Physique des Plasmas (LPP), 4 Place Jussieu, 75005 Paris, France
        \and
        SETI Institute, 339 Bernardo Ave, Mountain View, CA 94043, USA
	    \and 
	    Lockheed Martin Solar and Astrophysics Laboratory, 3251 Hanover Street, Palo Alto, CA 94306, USA
        }

    \date{Received 8 October 2025; accepted 10 December 2025}
 
    \abstract
    {
    The formation mechanism for the dynamic type~II spicules has remained elusive for many years. 
    Their dynamical behaviour has long been linked to magnetic reconnection, yet no conclusive evidence has been provided. 
    However, one recent observational study found signs of magnetic reconnection, as traced by Ellerman bombs (EBs), at the footpoints of many spicules. 
    The triggering of EBs is generally linked to magnetic reconnection due to flux emergence and convective motions in the photosphere.
    }
    {
    We aim to explore whether we can connect EBs to type~II spicules, and determine to what extent we can use EBs as an observational proxy to probe magnetic reconnection in this dynamic. 
    We also aim to provide further insight into the mechanisms that trigger EBs. 
    }
    {
    We used a simulation run with the radiative magnetohydrodynamics code Bifrost to track spicules and study the physical processes underlying their formation. 
    To detect EBs and classify the spicules, we synthesised the chromospheric $\Ha$ spectral line using the multilevel radiative transfer code RH1.5D. 
    We also traced shocks and current sheets to decipher the origin of EBs and spicules. 
    We selected one type~II spicule with a strong EB near its footpoint and studied their formation in detail.
    }
    {
    A magnetoacoustic shock advects the magnetic field lines towards an oppositely directed ambient field, creating a current sheet.
    The current sheet accelerates dense plasma via a whiplash effect generated by magnetic reconnection into the inclined ambient field, launching the spicule. 
    Several EB profiles trace shock- and magnetic-reconnection-induced dynamics during this process at the spicule footpoint.
    }
    {
    We present a new EB triggering mechanism in which a shock-induced current sheet reconnects, triggering an EB in the lower solar atmosphere.
    The shock-induced current sheet generates the upwards propagation of a type~II spicule via reconnection outflows.
    These results provide a plausible physical origin for the recently observed connection between EBs and spicules.
    }
    

    \maketitle

\section{Introduction}
\label{introduction}

    Spicules are thin, highly dynamic plasma jets that rise from the Sun's lower atmosphere.
    Depending on their observed dynamics, they are generally categorised as either type~I or type~II spicules \citep[see, e.g. the reviews by][]{1972ARA&A..10...73B, 2000SoPh..196...79S, 2012SSRv..169..181T}.
    When we observe spicules in chromospheric lines, for example, $\Ha$ or $\CaII$~H, type~I spicules show an apparent rise and fall motion, while type~II spicules rise quickly before rapidly fading out of wideband passbands due to heating \citep[][]{2007PASJ...59S.655D, 2014ApJ...792L..15P}.

    Type~II spicules are also more dynamic and ubiquitous, residing in the quiet Sun and coronal holes, whereas type~I spicules are mainly found in active regions \citep{2012ApJ...759...18P}.
    Type~I spicules have been established to be driven by magnetoacoustic shocks \citep{2006ApJ...647L..73H, 2007ApJ...655..624D, 2012ApJ...744L...5J}, while the driving mechanism of type~II spicules is still under debate.
    
    Radiative-magnetohydrodynamic (rMHD) simulations suggest that type~II spicules may be driven by a whiplash effect, inferred as the release of amplified magnetic tension \citep[][hereafter \citetalias{2017Sci...356.1269M}]{2017Sci...356.1269M}.
    However, the authors did not establish a causal connection between magnetic reconnection and the whiplash effect, leaving this underlying origin to be investigated.
    Alternatively, a resistive MHD model and several observational studies argue that type~II spicules may be driven by magnetic reconnection from the interaction between emerging and pre-existing magnetic fields \citep{2011ApJ...731L..18M, 2011A&A...535A..95D, 2016ApJ...828L...9S, 2019Sci...366..890S}.

    Another phenomenon linked to magnetic reconnection from the interaction between emerging and pre-existing magnetic fields is Ellerman bombs \citep[EBs;][]{1917ApJ....46..298E}.
    Ellerman bombs are short-lived, intense brightenings in regions with strong magnetic fields and are found throughout the solar disc, in both active regions and the quiet Sun \citep{2016A&A...592A.100R}.
    They are typically observed in the line wings of $\Ha$ as bright features in the photosphere.
    Their spectral signatures are distinguished from typical magnetic bright points by a moustache-like profile, i.e. strong emission in the line wings with an unaffected line core, and by their bright flickering dynamics and flame-like morphology when observed away from the disc centre \citep{2011ApJ...736...71W}.
    It is commonly accepted that EBs are possible tracers of magnetic reconnection in the photosphere in observations \citep{2002ApJ...575..506G, 2004ApJ...614.1099P, 2007A&A...473..279P, 2006ApJ...643.1325F, 2008ApJ...684..736W, 2025A&A...693A.221B, 2025A&A...698A.174B}, which is further supported by simulations \citep{2017A&A...601A.122D, 2017ApJ...839...22H}.

    Further support for EBs being tracers of magnetic reconnection comes from their association with another small-scale reconnection phenomenon in the lower atmosphere: UV bursts \citep{2018SSRv..214..120Y}. 
    UV bursts are detected by the Interface Region Imaging Spectrograph \citep[IRIS;][]{2014SoPh..289.2733D} as small, intense, and transient brightenings in the \ion{Si}{IV}~1394 and 1403 lines
    \citep{2014Sci...346C.315P}. 
    This phenomenon has been proposed to form in a wide range of temperatures \citep[35--100~kK;][]{2019A&A...626A..33H, 2019A&A...627A.101V, 2024A&A...685A...2C}, which, among other things, has raised the question of whether the plasma in EBs can be heated above $10^{4}$~K.
    Their connection to EBs was first speculated by \citet{2014Sci...346C.315P}, which has occasionally been confirmed in observations \citep{2015ApJ...812...11V, 2015ApJ...810...38K, 2016A&A...593A..32G, 2016ApJ...824...96T, 2019ApJ...875L..30C}, and later confirmed in a 3D rMHD simulation \citep{2019A&A...626A..33H}.

    The process behind the reconnection that triggers EBs has mainly been linked to undulating magnetic field lines and the interaction between emerging and pre-existing magnetic fields. 
    \citet{2002ApJ...575..506G}, \citet{2004ApJ...614.1099P, 2007A&A...473..279P}, and \citet{2008ApJ...684..736W} reported on how reconnection occurs along the separatrix in U~loops related to dipolar magnetic features \citep[also called bald patches, defined as where the magnetic field is tangent to the photosphere along a magnetic inversion line, defining the bottom of the U~loop;][]{1993A&A...276..564T}. 
    \citet{2002ApJ...575..506G} also reported a scenario where the reconnection occurs along a quasi-separatrix layer between a vertical emerging field and pre-existing horizontal field, while \citet{2008ApJ...684..736W} observed EBs located at the top of an $\Omega$~loop. 
    The simulation work of \citet{2017A&A...601A.122D} and \citet{2017ApJ...839...22H} illustrated how the reconnection occurred as the oppositely directed magnetic footpoints of one or two $\Omega$~loops moved into each other. 
    The general conception in these studies is that the reconnection is related to magnetic flux emergence and convective motions.
    
    The hypothesis of magnetic reconnection as a driving mechanism for type~II spicules was recently well supported in \citet[][hereafter \citetalias{2025A&A...697A.180S}]{2025A&A...697A.180S}.
    In a 40-min time series of the quiet Sun, we found nearly 600 cases, 80 of which were very suggestive, of individual connections between photospheric magnetic reconnection and type~II spicule formation.
    We inferred magnetic reconnection in the photosphere from the detection of EBs and occasional flux cancellation near the apparent footpoints of the spicules. 
    However, a causal link between reconnection and spicules remains debated, and details of the connection between EBs and spicule dynamics remain unclarified and require further investigation.

    To further investigate this proposed connection, which was also shown by \citet{2025A&A...698A.174B}, we examined a numerical simulation that shows self-consistent spicule generation \citep[][hereafter \citetalias{2020ApJ...889...95M}]{2020ApJ...889...95M}.
    This simulation is an extension of the simulation presented in \citetalias{2017Sci...356.1269M}. 
    We chose a type~II spicule event with a strong EB at its footpoint to examine the triggering mechanisms behind them. 
    We also analyse the relation between EBs and the formation of type~II spicules, as well as the resulting synthetic emissions.

    Ultimately, we provide novel support and evidence addressing three long-standing questions in the field: whether the well-established EB triggering processes are the only mechanisms capable of producing EBs; whether magnetic reconnection can finally be linked to type II spicules; and whether the observed connections between EBs and type II spicule formation represent causal relationships.

\section{Methods}
\label{sect: methods}
    
    \subsection{Numerical simulation}
    \label{meth: simulation}
    
        To inspect the observed connection between EBs and spicule formation, we used a 2.5D simulation of two plage-like regions with opposite polarities and self-consistent type~II spicule generation. 
        This simulation was created with the state-of-the-art rMHD code, Bifrost \citep{2011A&A...531A.154G} and is presented by \citetalias{2020ApJ...889...95M}.
        It is an extension to \citetalias{2017Sci...356.1269M}'s simulation, which uses a version of the generalised Ohm's law, including ambipolar diffusion and the Hall term in the induction equation. 
        \citetalias{2020ApJ...889...95M} also includes non-equilibrium ionisation (NEI) of both hydrogen and helium, which can lead to a greater temperature increase at heating sites, such as shocks and regions with large currents in type~II spicules. 
        More details on the simulation are given in Appendix~\ref{subapp: sim_conf}.

    \subsection{Synthetic spectra}
    \label{meth: radiative_transfer}

        To calculate synthetic spectra for the spicule and its footpoint, we used the state-of-the-art radiative transfer code, RH 1.5D \citep{2015A&A...574A...3P, 2001ApJ...557..389U}\footnote{https://github.com/ITA-Solar/rh}. 
        RH 1.5D is a multilevel, non local thermodynamic equilibrium (NLTE) radiative transfer code that allows partial frequency redistribution. 
        It solves the NLTE radiative transfer for 3D and 2D simulations, treating every column of the atmosphere in a plane-parallel approximation (1.5D). 
        We calculated the synthetic spectra using a 6-level hydrogen atom model, i.e. five levels plus continuum.

        To get synthetic spectra that are more suitable for comparison with observations, we smeared the synthetic profiles with a representative instrumental point spread function (PSF). 
        We smeared the synthetic spectra by convolving them with the PSF of the CRISP instrument \citep{2008ApJ...689L..69S} at the Swedish 1-m Solar Telescope \citep[SST;][]{2003SPIE.4853..341S}, which has been used extensively for studies of EBs and spicules in $\Ha$ \citep[e.g. ][]{2009ApJ...705..272R, 2011ApJ...736...71W}.

    \subsection{Tracking shocks and magnetic reconnection}
    \label{meth: tracking_physics}

        Ellerman bombs are associated with localised heating events in the lower solar atmosphere \citep{2002ApJ...575..506G}, where the density is relatively high. 
        In such plasma regimes characterised by high Reynolds and Lundquist numbers, strong gradients in velocity $\mathbf{v}$ and magnetic field $\mathbf{B}$ are expected to drive enhanced energy dissipation and heating, in the form of shocks and current sheets.
        Following the detection methodology presented by \citet{2025arXiv251101858N}, we identify shocks and reconnecting current sheets using the criteria 
        \begin{equation}
            -\mathbf{\nabla}\cdot\mathbf{v}>\frac{c_s}{\epsilon\cdot ds},
            \label{eq: alpha_crit}
        \end{equation}
        \begin{equation} 
            \frac{|\mathbf{\nabla} \times \mathbf{B} \cdot \mathbf{B}|}{\mathbf{B}^2} > \frac{1}{\epsilon\cdot ds}, 
            \label{eq: cs_crit} 
        \end{equation}
        respectively, where $c_s$ is the sound speed, $ds={\rm max}(dx,dz)$ is the largest grid spacing, and $\epsilon=6$ is a dimensionless threshold tuned to distinguish nonlinear dissipation (e.g. shocks and current sheets) from linear wave propagation and broader dissipative structures. 

        The normalised parallel current (Eq.~\ref{eq: cs_crit}) can be regarded as a restrictive form of the common parameter $\alpha = \mathbf{J} / \mathbf{B}$. 
        This formulation enables us to target the core regions of current sheets, where reconnection occurs \citep[see also][]{hesseTheoreticalFoundationGeneral1988,schindlerGeneralMagneticReconnection1988}.

\section{Results}
\label{sect: results}

\subsection{Selection of the EB and spicule pair}
\label{res: selection_EB_spicule}

    To look for EBs connected to spicule formation, we performed $\Ha$ spectral line synthesis on all columns that captured spicules and their footpoints, including an additional 0.5~Mm on each side.
    We found several events of EBs forming in the area of the spicules' footpoints just before the spicules started to form.
    The event selected for an in-depth case study had the strongest EB profile, i.e. the strongest emission in the $\Ha$ wings at the spicule's footpoint.

    \subsection{Classification of the spicule}
    \label{res: the_spicule}
    
\begin{figure}
        \resizebox{\hsize}{!}{\includegraphics{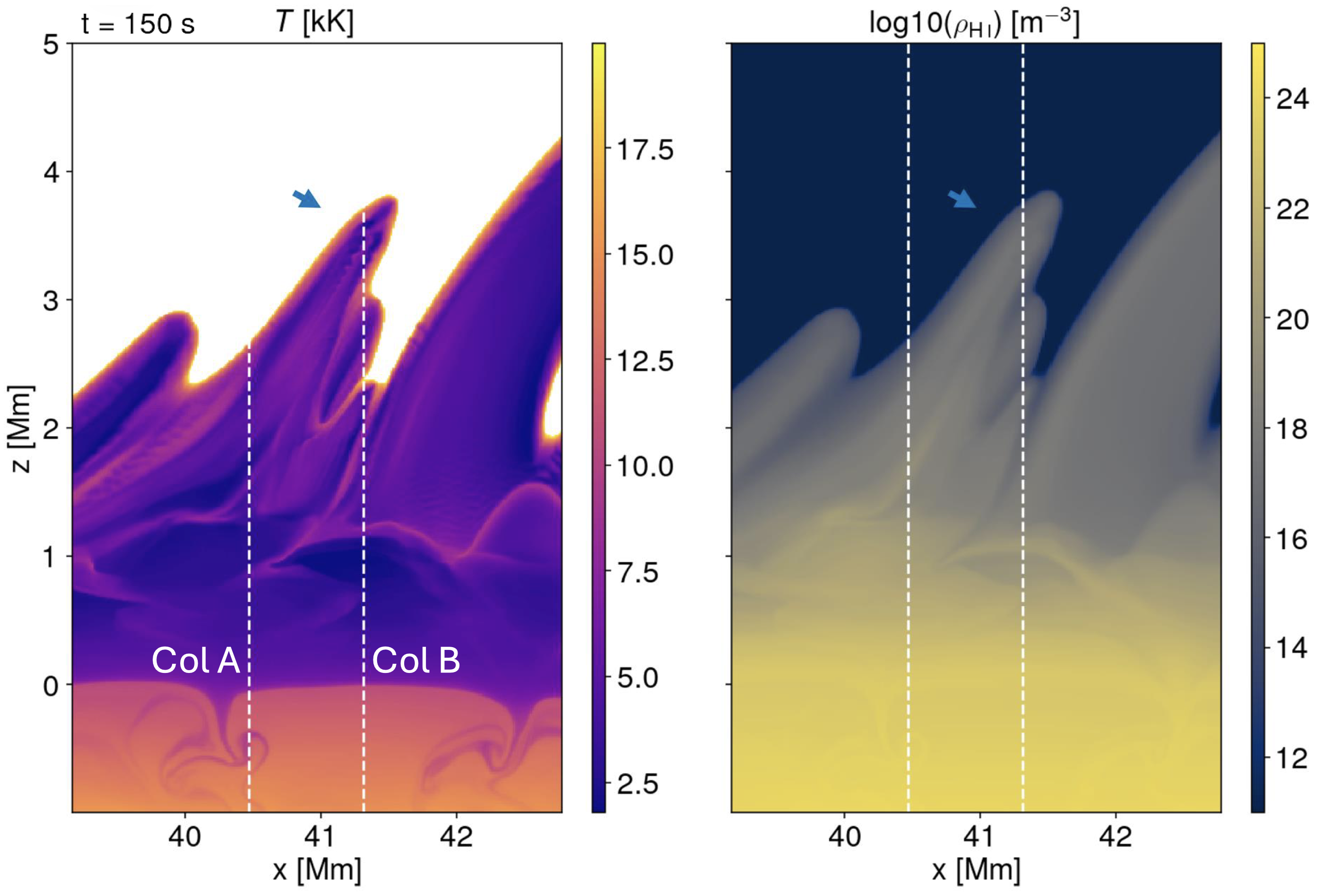}}
        \caption{
            Spicule at the beginning of its ascending phase, indicated by the blue arrows. 
            The left frame shows the temperature, saturated at 20~kK. 
            The right frame shows the logarithm of the neutral hydrogen number density. 
            The dashed lines represent the columns A and B used for calculating the $\Ha$ $\lambda t$~diagrams, presented in Fig.~\ref{fig: lambda_t}. 
            An animation of this figure is available at \url{http://tsih3.uio.no/lapalma/subl/recon_eb_spic/the_spicule_temp_dens.mp4}.
        }
        \label{fig: spicule}
        \end{figure}

        Our investigations focus on the spicule presented in Fig.~\ref{fig: spicule} and the corresponding animation. 
        This spicule is particularly interesting, showing a type~II signature and strong EB emissions at its footpoint.
        We measure the time $t$ relative to $t = 0$~s when the EB begins, which occurs at time step 290~s in the simulation; the spicule becomes visible in the synthetic spectra at $t = 110$~s after the onset of the EB. 

        The time series animation of temperature and density, associated with Fig.~\ref{fig: spicule}, shows distinct phases of rise and fall of the spicule. 
        However, in the temporal evolution of the $\Ha$ profiles, the falling phase and parts of the rising phase of the spicule plasma are not visible. 
        This fading in $\Ha$ is a key observational property of type~II spicules \citep{2009ApJ...705..272R}.

\begin{figure}
        \resizebox{\hsize}{!}{\includegraphics{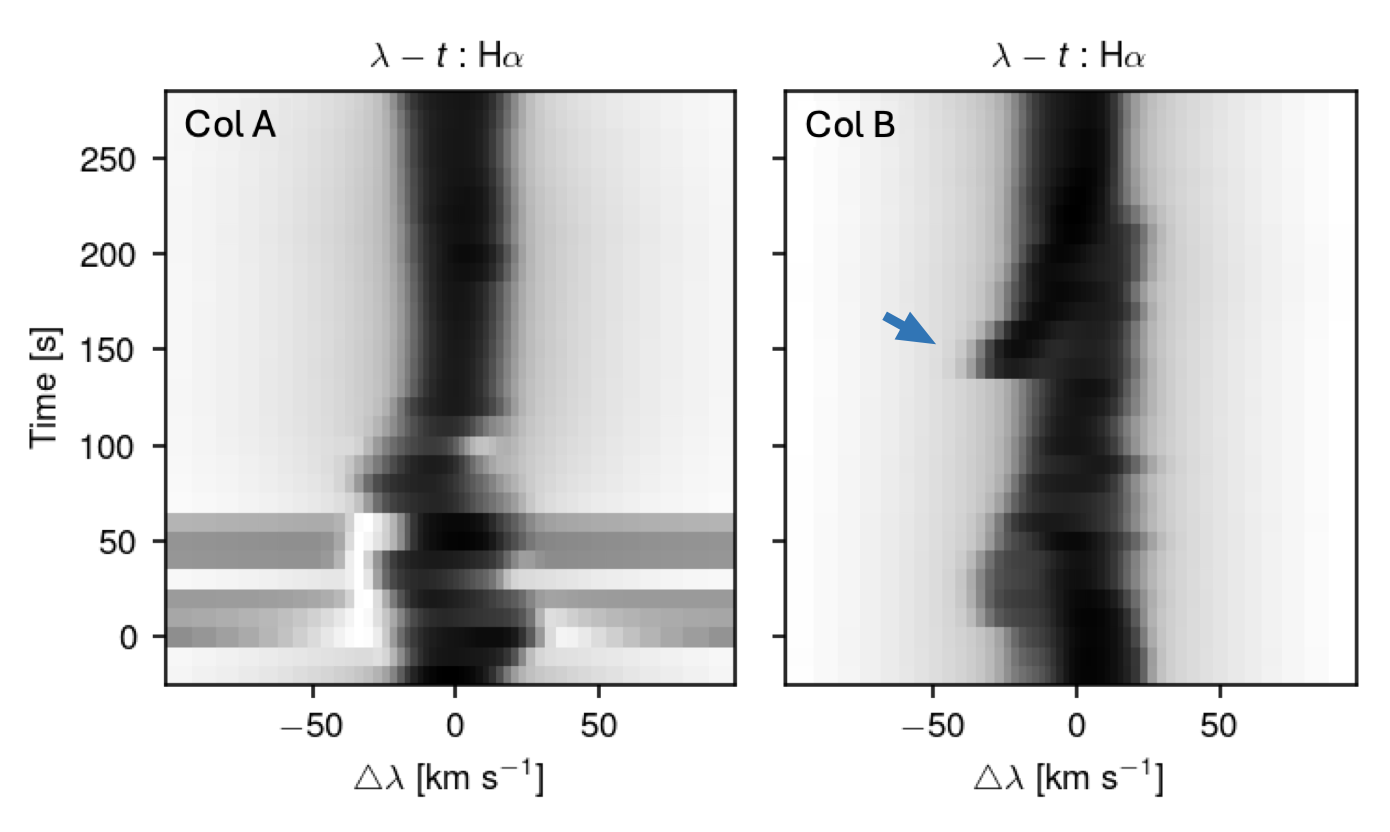}}
        \caption{
            Temporal evolution of the $\Ha$ synthetic spectral profiles in $\lambda t$ diagrams for the EB (left panel) and the connected spicule (right panel). 
            The EB and spicule spectra are calculated along the columns A and B marked by the dashed lines in Fig.~\ref{fig: spicule}. 
            The blue arrow indicates the spectral profile corresponding to the timeframe presented in Fig.~\ref{fig: spicule}. 
            The EB's beginning is marked by time $t = 0$, corresponding to 290~s in the simulation of \citetalias{2020ApJ...889...95M}.
        }
        \label{fig: lambda_t}
        \end{figure}

        To demonstrate the fading of the spicule, we present the $\Ha$ spectral evolution in the form of a $\lambda t$~diagram for column B in Fig.~\ref{fig: lambda_t}.
        As the spicule rises, the $\Ha$ spectral line is blueshifted, as expected from an upwards propagation along the line of sight and as shown by the excursion indicated by the blue arrow. 
        At $\sim$40~s after the spicule is visible in the right $\lambda t$~diagram, the blueshifted depression of the spectrum is absent even though the spicule continues to rise rapidly, giving a sudden excursion in the blue wing of the $\lambda t$~diagram. 
        This sudden excursion in the blue wing and the absence of the transition to a subsequent redshift demonstrate that the spicule is no longer visible in $\Ha$ in the later stages of its evolution when the plasma falls back down. 

        In observations, this corresponds to the spicule fading out of the instrumental bandpass filter. 
        This behaviour in $\Ha$ $\lambda t$~diagrams resulted in the term rapid blueshifted excursions \citep[RBEs;][]{2008ApJ...679L.167L, 2009ApJ...705..272R} often used for the on-disk counterpart of type~II spicules. 
        To demonstrate that the fading of the spicule in $\Ha$ is visible in all the columns that capture the spicule, a series of $\lambda x$~diagrams is shown in Appendix \ref{subapp: type II}, Fig.~\ref{fig: lambda_x}.

        \begin{figure}
        \resizebox{\hsize}{!}{\includegraphics{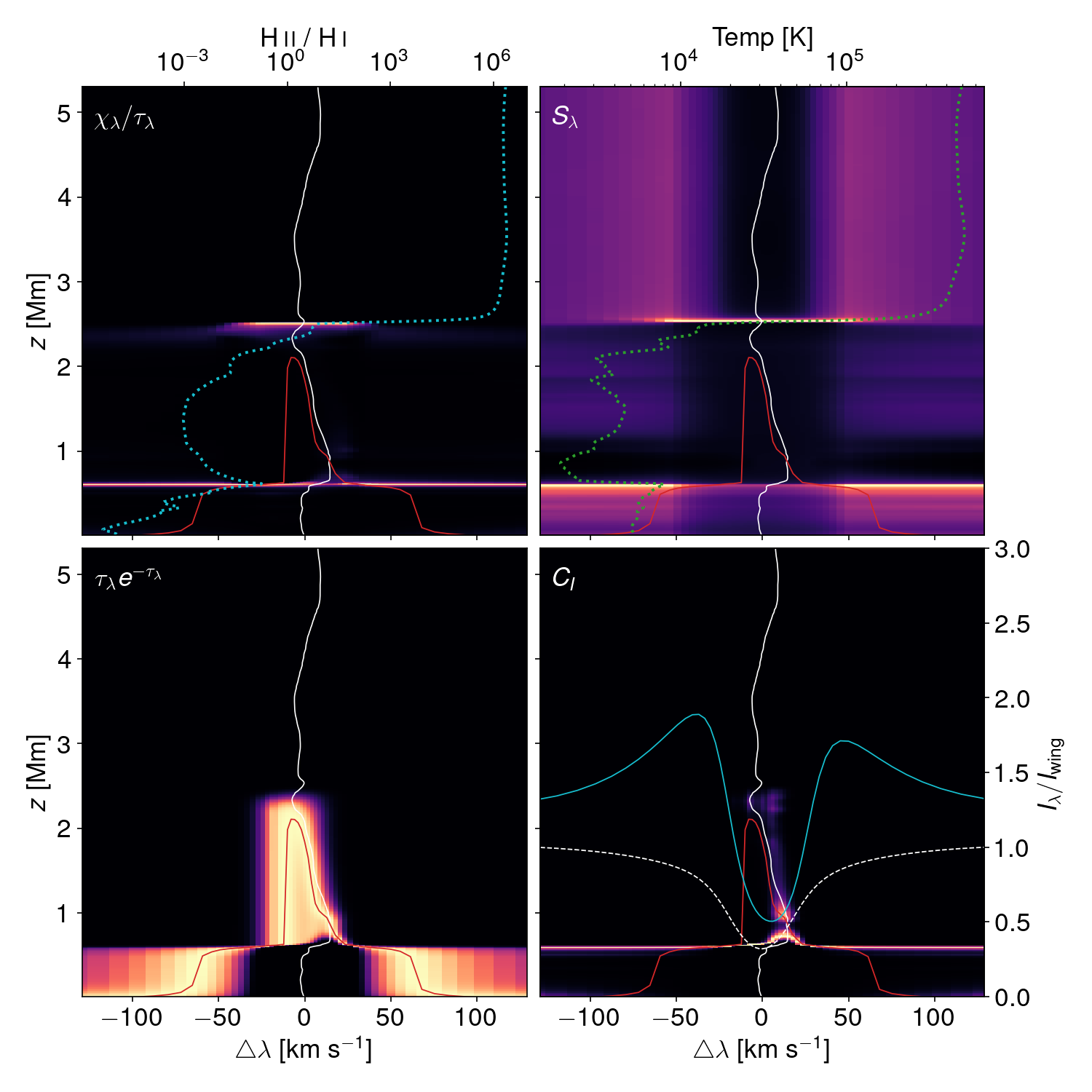}}
        \caption{
            Formation of the $\Ha$ intensity profile for the EB at $t=0$ from column A marked in Fig.~\ref{fig: spicule}. 
            The contribution function to emergent intensity $C_I$ in the lower right panel can be calculated as the product of the three factors shown in the other panels. 
            The height where $\tau_{\lambda} = 1$ (solid red) and the velocity profile in the $z$-direction (solid white) are plotted in all panels (negative velocities correspond to blueshift, i.e. upflows, while positive values are downflows). 
            The top left panel shows the factor $\chi_{\lambda} / \tau_{\lambda}$, i.e. the ratio between the opacity and optical depth; the dotted cyan line shows the ionised to neutral hydrogen ratio as a function of height. 
            The top right panel shows the monochromatic source function $S_\lambda$; the dotted green line shows the atmosphere's temperature profile. 
            The lower left panel shows the factor $\tau_{\lambda} / e^{-\tau_{\lambda}}$, i.e. a function that peaks in the region where $\tau_{\lambda} \approx 1$. 
            The lower right panel shows the contribution to emergent intensity $C_I$. 
            The solid cyan line shows the $\Ha$ spectral profile of the EB, while the dotted white line shows an $\Ha$ background profile, i.e. the average profile over many columns and scans. 
            The spectral profile is normalised to $I_{\text{wing}}$, which is the average intensity of the outermost wavelength points in the blue and red wings of the background profile.
        }
        \label{fig: cont_func}
        \end{figure}

    \subsection{The location of the Ellerman bomb}
    \label{res: contribution_function}

        \begin{figure*}
            \resizebox{\hsize}{!}{\includegraphics{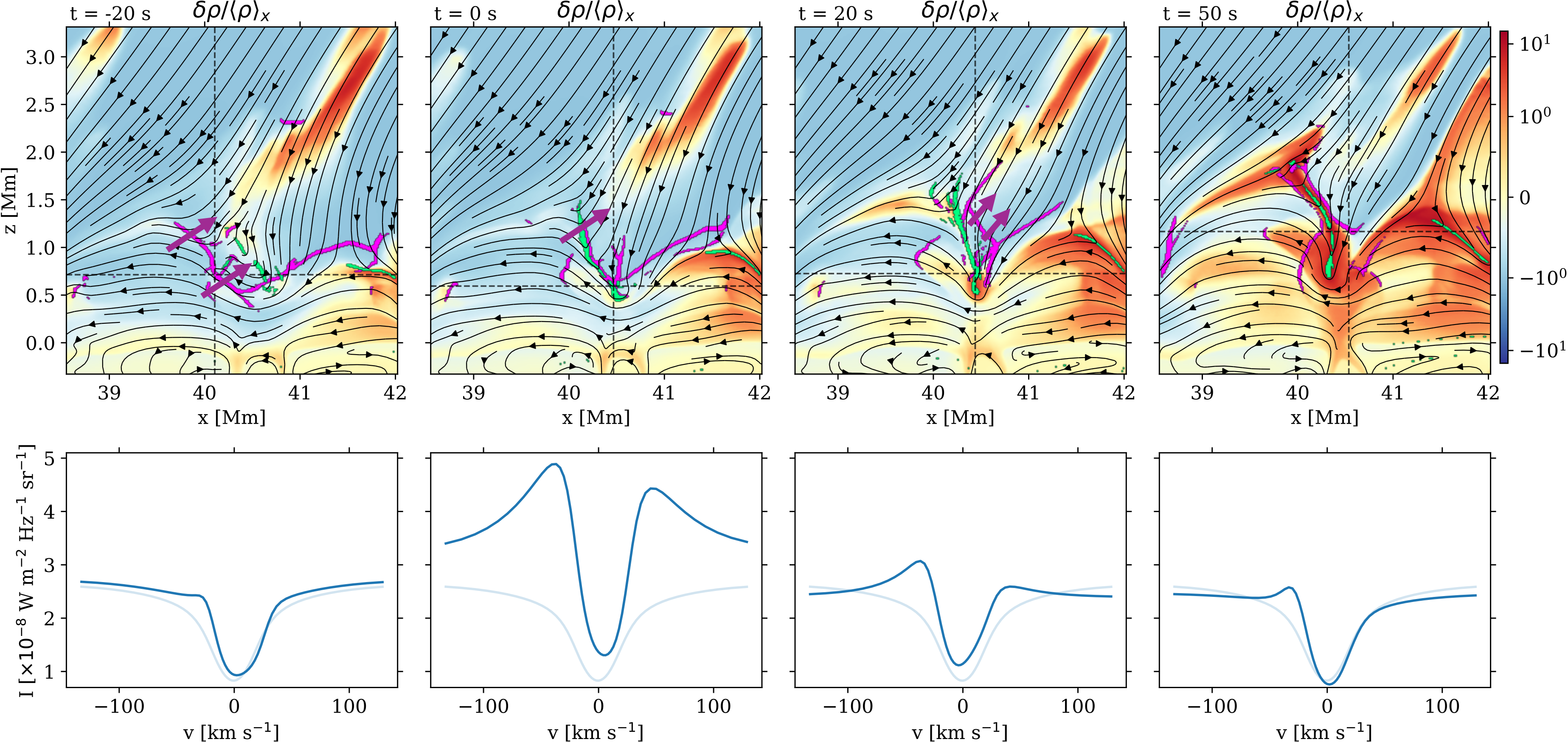}}
            \caption{
                Time series illustrating the four main physical processes that trigger the EB and form the spicule. 
                From left to right: 
                propagation of a magnetoacoustic shock front;
                the generation of a current sheet by the shock;
                formation of a reconnecting current sheet;
                plasma ejection via reconnection outflows, ultimately leading to the formation of the rising spicule. 
                \textit{Top}: Normalised density fluctuation maps, $\delta\rho/\langle\rho\rangle_{x}$, with detected shocks and current sheets overlaid in magenta and green, respectively. 
                Thick purple arrows indicate the direction of the shock's motion. 
                The solid lines with arrows indicate the direction of the magnetic field. 
                The vertical, dotted line marks the column that is used for integration of the synthetic spectrum; 
                the horizontal, dotted line marks the formation height of the EB; 
                the intersection between the dotted lines marks the location of the EB. 
                \textit{Bottom}: EB spectra (strong colour) corresponding to the column marked by the vertical dotted line, along with an $\Ha$ background profile (weak colour), i.e. the average profile over many columns and snapshots. 
                An animation of the upper panels is available at \url{http://tsih3.uio.no/lapalma/subl/recon_eb_spic/l2d90x40r_it_eb5_ShCsDyn2D5_320_355_1_5fps.mp4}.
            }
            \label{fig: physics}
        \end{figure*}

        \begin{figure*}
            \resizebox{\hsize}{!}{\includegraphics{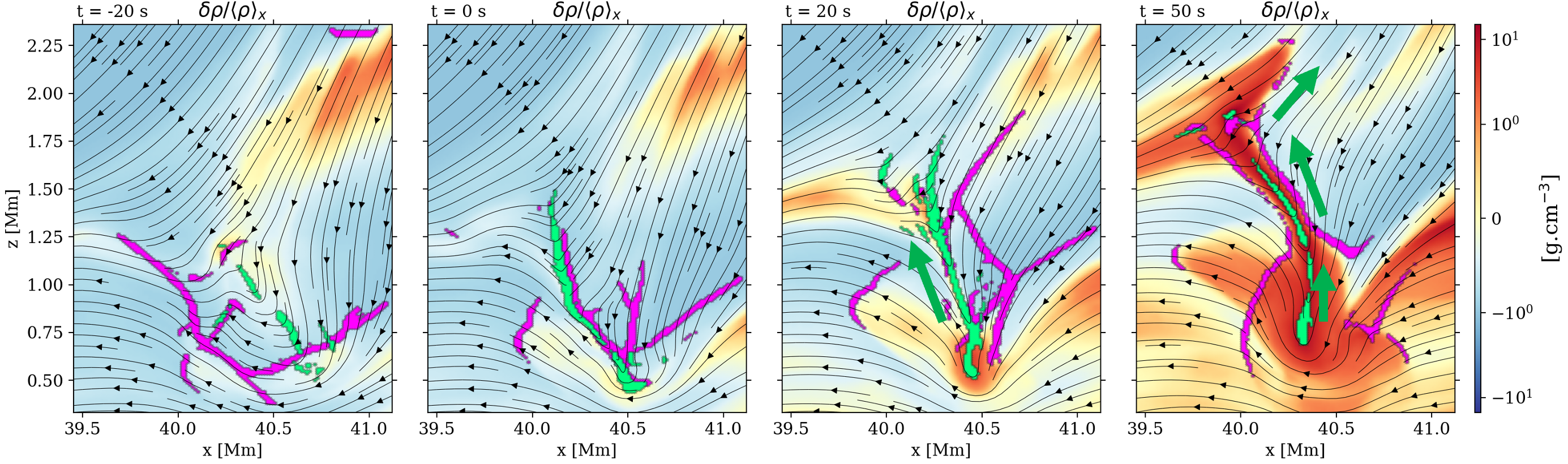}}
            \caption{
                Close-up of the mechanisms that trigger the EB and form the spicule. 
                Thick green arrows indicate key flow directions for the plasma related to the current sheet (more details on the plasma flow directions are presented in Fig.~\ref{fig: Quiver}). 
                Everything else follows the convention of Fig. 4.
            }
            \label{fig: zoomed_physics}
        \end{figure*}

        The event was selected because it contained the strongest EB. 
        The strong EB wing emission is visible in the $\lambda t$~diagram for column A in Fig.~\ref{fig: lambda_t}, which starts at $t=0$~s. 
        To establish a connection between the EB and the spicule, we first determined the height of the origin of the enhanced wing emission.

        We determined the height using the 4-panel analysis of the contribution function described by \citet{Carlsson_1997}.
        This function describes the contribution to the emergent intensity from a source as a function of wavelength and height (more details are given in Appendix~\ref{app: cont_func}).
        By investigating the contribution function presented in the lower right panel of Fig.~\ref{fig: cont_func} at wavelengths corresponding to the EB emission features, we can pinpoint the geometrical height of the EB at the spicule footpoint.

        The contribution function shows that the main contribution to the EB emission peaks originates from a narrow height interval, just above 0.5~Mm. 
        At this height, looking at the upper panels of Fig.~\ref{fig: cont_func}, we also see a sharp peak in both the ratio of ionised to neutral hydrogen (top left; peak value is 0.176) and temperature (top right; peak value is 7984~K), demonstrating an energetic heating event. 
        To pinpoint the height of the EB, we chose the height where the contribution function in the lower right panel is highest at the wavelength corresponding to the strongest emission peak.
        The height of the EB in Fig.~\ref{fig: cont_func} is 0.59~Mm.

    \subsection{Triggering of the Ellerman bomb}
    \label{res: eb_trigger}

        To investigate the triggering of the EB, we present details of four distinctive dynamical phases in Fig.~\ref{fig: physics}, with associated EB profiles. 
        Before the onset of the EB, acoustic waves in the photosphere steepen their fronts as they propagate upwards due to the sudden decrease in density and temperature with height. 
        Then, 20~s before the onset of the EB, the steepening of the wave front has led to a magnetoacoustic shock front according to the detection criterion of Eq.~\ref{eq: alpha_crit} and as shown by the magenta contours in the first panel.

        The initial shock of the event propagates upward, as indicated by the purple arrows and illustrated in the associated animation, towards a broad current layer along a U~loop formed by atmospheric motions. 
        This current layer is stable and contains smaller patches that are captured by the current sheet criterion of Eq.~\ref{eq: cs_crit}, as shown by the green contours. 
        We note a subtle enhancement in the blue wing of the spectrum, hinting towards the EB about to be triggered.

        As the magnetoacoustic shock propagates through the atmosphere, it drags and bends the magnetic field lines in the high plasma~$\beta$ environment, where $\beta$ describes the ratio of the thermal plasma pressure to the magnetic pressure. 
        The current layer is compressed into a thin, long current sheet as the shock arrives at the pre-existing U~loop, according to the current sheet detection criterion. 
        This is consistent with the substantial bending of the magnetic field line along the current sheet, implying small-scale dynamics.
        
        The current sheet is co-located and coupled to the shock, triggering the formation of an EB inside it. 
        The location of the EB is pinpointed by the intersection between the vertical and horizontal dotted lines. 
        The vertical line marks the column over which the spectrum is integrated, serving as the line of sight, while the horizontal line is the height of the EB as determined from the contribution function. 
        The shock's propagation advects the magnetic field lines and the current sheet towards an oppositely directed magnetic field, as indicated by the purple arrow.

        Then, 20~s after the onset of the EB, the advected field lines interact with the ambient magnetic field, and the current sheet is formed and no longer coupled to the shock, as shown in the third panel. 
        The EB is located inside the now reconnecting current sheet, as illustrated by the dotted lines. 
        The shock has decoupled from the current sheet and continues to propagate up and to the right, as indicated by the purple arrows in the third panel.

        Towards the end of the EB's lifetime (last panel), two shocks form a corridor-like structure around the current sheet. 
        The EB is no longer coupled to the reconnecting current sheet, but is now located inside the shock above the current sheet. 
        We note that the Lorentz force work is positive at this location, indicating conversion from magnetic to kinetic energy and suggesting a slow-mode nature for the shocks outlining the shock corridor.
        
    \subsection{The spicule's formation mechanism}
    \label{res: spic_driver}

    \begin{figure*}
        \begin{center}
            \includegraphics[width=\linewidth]{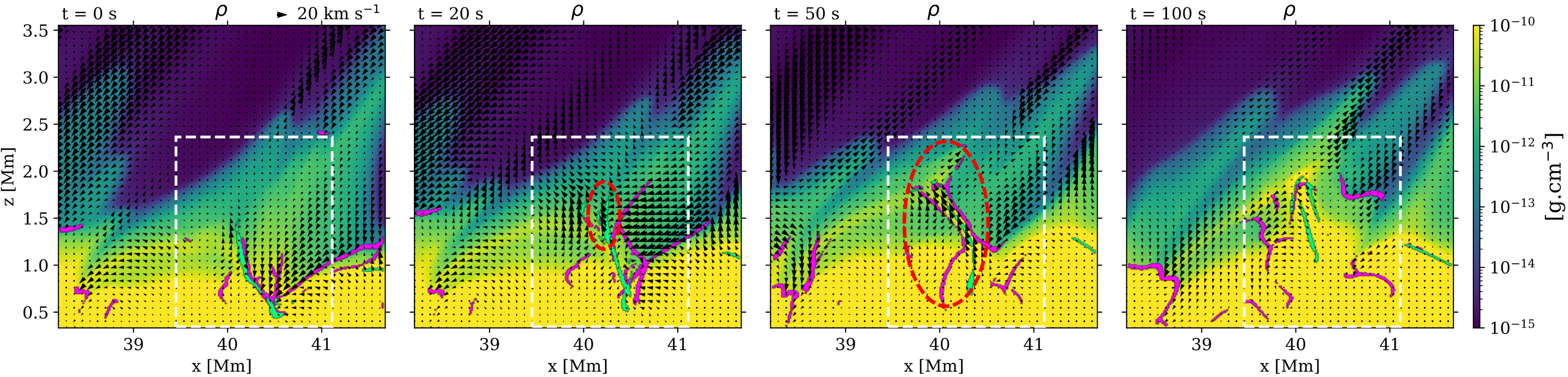}
        \end{center}
        \caption{
        Time series illustrating how the velocity field is impacted by the shock-current-sheet-reconnection chain of events presented in Fig.~\ref{fig: physics}. 
        The background shows the density field on a logarithmic scale, with overplotted regions where either shocks (magenta) or current sheets (green) are detected, along with the velocity field shown with dark arrows. 
        An indicative 20~$\kms$ arrow above the left panel is a reference for the scale. 
        The white dashed rectangle outlines the field of view for the upper panels in Fig.~\ref{fig: physics}. 
        The red dashed ellipse in the third panel highlights the area where the reconnection jet forms. 
        An animation of this figure is available at \url{http://tsih3.uio.no/lapalma/subl/recon_eb_spic/l2d90x40r_it_eb5_ShCsQuiv2D5_320_355_1_5fps.mp4}.
        }\label{fig: Quiver}
    \end{figure*}

        To investigate the mechanisms behind the formation of the spicule, we present a zoomed-in version of Fig.~\ref{fig: physics} in Fig.~\ref{fig: zoomed_physics}. 
        The previous subsection showed how a magnetoacoustic shock compressed a thick current layer into a thin, reconnecting current sheet by advecting field lines towards an oppositely directed ambient field, triggering the EB at the spicule's footpoint. 
        The mechanisms that triggered the EB are the same as those that subsequently trigger the spicule.

        When the reconnecting current sheet forms 20~s after the onset of the EB (third panel), the field lines along it are bent by almost 180$\degr$.
        This extreme refraction of the field lines amplifies the magnetic tension, which is released when the advected field lines reconnect with the ambient magnetic field, confirmed by the detection criterion of Eq.~\ref{eq: cs_crit}. 
        The whiplash effect of the reconnection slingshots the plasma upward, as indicated by the green arrow in the third panel, with speeds up to 23 $\kms$.

        Then, 30~s later (fourth panel), reconnection outflows accelerate dense plasma from below and eject the spicule along the magnetic field lines via reconnection outflows, as indicated by the green arrows. 
        As dense plasma is propelled along the current sheet, a shock corridor forms around it, according to the detection criterion of Eq.~\ref{eq: alpha_crit} and shown by the magenta contours. 
        The ejected plasma represents the beginning of the spicule, which can be seen at $(x,~z) = (40.1,~2.1)$~Mm.
        
        We finally note a termination shock at (39.9,~1.6) Mm, as the reconnecting plasma crashes into the perpendicular ambient field with velocities up to 26 $\kms$.
        To present the Mach number of the plasma, we note that the Alfvén speed and the sound speed in this area just before the reconnection jets are affecting these parameters are around 15 and 10 $\kms$, which gives a fast-mode speed of $\sqrt{15^2 + 10^2} = 18\ \kms$.
        This gives the plasma propelled along the current sheet a Mach number of around 1.4.
        
        To establish the relation between the upflow driven by reconnection and the subsequent formation of the spicule, we illustrate the velocity field in Fig.~\ref{fig: Quiver} along the shock-current-sheet-reconnection chain of events (left, second, and third panels, respectively), as well as a later time after the onset of the spicule (right panel). 
        We confirm that strong flows (of the order of $20$~$\kms$ and stronger) are created in the current sheet region, i.e. the area outlined by the dashed red ellipse in the second and third panels. 
        This shows that plasma starts to rise at the tip of the corridor towards the upper right, which later initiates the spicule's rise (right panel).

\section{Discussion}
\label{sect: discussion}

    This work was motivated by the observational findings of \citetalias{2025A&A...697A.180S}, which suggested a possible causal link between type~II spicules and EBs at their footpoints. 
    Here we explore the formation mechanisms that trigger a spicule and an EB-like brightening at its footpoint in the rMHD simulation by \citetalias{2020ApJ...889...95M}, and the connection between them. 
    We investigated this by tracking magnetoacoustic shocks and current sheets using the criteria presented in Sect.~\ref{meth: tracking_physics}.

    During our investigations, we found answers to the three research questions in the Introduction: 
    We discovered a new process that can trigger reconnection and form an EB. 
    This process is unrelated to the conventional drivers of EB formation, such as flux emergence and magnetic footpoint motions. 
    We further disclosed the elusive link between magnetic reconnection and spicule formation. 
    We discuss these answers in detail below.

    \subsection{Triggering of the Ellerman bomb}
    \label{disc: eb_trigger}

        We showed in Sect.~\ref{res: eb_trigger} how a magnetoacoustic shock compressed a thick current layer into a thin, reconnecting current sheet and triggered the EB. 
        This is similar to what \citet{2021ApJ...922..219N} reported when they examined multiple current sheets that trigger the tearing instability after being compressed by a passing shock wave. 
        In our case, we have a thick current layer that is compressed into a thin current sheet by the shock, and later reconnects as the magnetic field lines are advected towards the opposite-directed ambient field, presumably due to the nature of the shock.
        
        The shock shows geometric evidence of a fast-mode shock, as the magnetic field lines in some places are refracted away from the shock normal 
        \citep[see also Fig.~5.6 of][]{2014masu.book.....P}.
        The geometry of a fast-mode shock is more suggestive at the onset of the EB, i.e. at $t = 0$~s, where the shock's obliquity is further seen with the change of the magnetic field line orientation through the shock front.
        The heavily refracted field lines also appear to be enhanced by the advection of the field lines towards the oppositely directed ambient field.

        Furthermore, the departure of the initial shock from the current sheet may arise from the conversion from a fast- to a slow-mode shock as $\beta$ decreases. 
        In the high-$\beta$ regime, the fast-mode shock drags the magnetic field. 
        Upon reaching a lower $\beta$, it transitions to a slow-mode shock that follows the field lines rather than bending them. 
        This interpretation is consistent with the plasma $\beta$ being above 15 at the current sheet location at $t = 0$~s, while the departing shock shows $\beta \approx 3$ along its front at $t = 20$~s. 
        Nevertheless, a dedicated, energetic analysis would be required to confirm the shock's nature and/or conversion, which is beyond the scope of our study.

        Regardless of the shock's nature, we report a new driver for magnetic reconnection in the lower atmosphere that can trigger an EB. 
        Conventional processes triggering EBs have been reported to be due to magnetic reconnection \citep{2002ApJ...575..506G, 2004ApJ...614.1099P, 2007A&A...473..279P, 2006ApJ...643.1325F, 2008ApJ...684..736W, 2017A&A...601A.122D, 2017ApJ...839...22H}, which are interpreted to be driven by flux emergence and magnetic footpoint motions. 
        Here, we have shown shock-induced, smaller-scale dynamics: 
        An upward-propagating shock drags the field lines and compresses a thick current layer into a thin, reconnecting current sheet, triggering the EB. 
        To further demonstrate that the event analysed here is unrelated to flux emergence, supplementary evidence is provided in Appendix~\ref{app: flux_emer}.

        A next step may be to investigate how frequently this chain of events contributes to the occurrence of EBs on the Sun, which is beyond the scope of this work. 
        We note that most studies examining EB drivers focus on active regions, where flux-emergence rates are about an order of magnitude higher than in the quiet Sun \citep{1985SoPh...97...51L}. 
        Although this difference is relevant for EBs driven by flux emergence, the shock-induced mechanism reported here does not rely on flux emergence. 
        Future work may therefore also consider how the differences in magnetic field strength and topology between active regions and the quiet Sun may affect the shock-induced chain of events.

    \subsection{Formation mechanism of the spicule}
    \label{disc: spic_driver}
    
        Similarly, as for the process of triggering EBs, the standard idea of the trigger for magnetic reconnection leading to spicule formation is the interaction between emerging and pre-existing magnetic fields \citep{2011ApJ...731L..18M, 2011A&A...535A..95D, 2016ApJ...828L...9S, 2019Sci...366..890S}. 
        However, linking magnetic reconnection to spicule formation is significantly more challenging than connecting it to EBs, due to the spatial separation between the inferred reconnection site and the forming spicule. 
        This has left the link between magnetic reconnection and spicule formation highly elusive. 
        Furthermore, our analyses of Figs.~\ref{fig: physics} and \ref{fig: zoomed_physics} show that flux emergence from below the photosphere is unnecessary to generate spicules, since small-scale dynamics can form a current sheet that reconnects and releases kinetic energy at the root of the subsequently rising spicules.

        This analysis is supported by the velocity plot in Fig.~\ref{fig: Quiver} and is consistent with the scenario proposed by \citetalias{2017Sci...356.1269M}, where type~II spicules result from a whiplash effect due to the release of magnetic tension. 
        Our analysis of Sect.~\ref{res: spic_driver} shows that this tension release is caused by magnetic reconnection from the interaction between oppositely directed magnetic fields along the current sheet, which is confirmed by the criterion in Eq.~\ref{eq: cs_crit}. 

        We note that the reconnection jet does not show typical bipolar flows. 
        This is expected, as we are close to the photosphere, where density increases significantly at lower altitudes.
        The density at the bottom of the current sheet is, on average, 75~times higher than that in the upper part and increases further down, serving as a barrier to reconnection downflows. 
        Additionally, and more importantly, the U~loop geometry configuration of the whole current-sheet structure preferentially directs the acceleration from the magnetic tension upward.

        Ellerman bombs have also been connected to UV bursts in both observations \citep{2015ApJ...812...11V, 2015ApJ...810...38K, 2016A&A...593A..32G, 2016ApJ...824...96T, 2018SSRv..214..120Y} and in a 3D rMHD simulation \citep{2019A&A...626A..33H}.
        As the EB in this work is confined to the lower part of the current sheet, it may be worthwhile to investigate whether the EB may be connected to a UV burst or a smaller UV brightening \citep[as in][and \citet{2024A&A...689A.156B}]{2017ApJ...845...16N} both observed by IRIS, in the upper part of the current sheet.

    \subsection{The Ellerman bomb’s connection to the spicule}
    \label{disc: eb_spic}

        From the observations presented in \citetalias{2025A&A...697A.180S}, we found 80 clear cases and over 500 more ambiguous cases linking EBs to spicule formation. 
        However, given the limitations inherent to an observational study, especially given the opacity gap observed in the Balmer lines \citep{2006A&A...449.1209L, 2012ApJ...749..136L}, we could not firmly conclude whether these connections were causal or coincidental.
        From our numerical simulations, we report a scenario of small-scale dynamics that triggers the formation of an EB and a spicule, thereby causally connecting the EB to the spicule.

        It is clear that the general correlation between EBs and spicule formation still needs further investigation: 
        In \citetalias{2025A&A...697A.180S} we observed many EBs without spicules connected to them, and most spicules did not show any EB signal at their footpoints.
        This can be attributed to different possible reasons, which are discussed in detail in \citetalias{2025A&A...697A.180S}. 
        
        One possible reason is also relevant for the analysis of our numerical simulation: 
        The wings of $\Ha$ are sensitive only to a certain height range in the atmosphere. 
        This means that there may be cases in our simulations with shock-induced reconnection leading to spicule formation, but without EB-signal in $\Ha$ at the footpoint of the spicule, simply because the reconnection happens outside the region the $\Ha$ wings are sensitive to. 

        Nevertheless, in Sect.~\ref{res: eb_trigger}, we showed how a magnetoacoustic shock formed a reconnecting current sheet that triggered an EB. 
        Then, we showed how the same chain of events launched the spicule through reconnection outflows from the reconnecting current sheet in Sect.~\ref{res: spic_driver}. 
        The combined analysis of the two ultimately causally connects the EB to the spicule formation.

        We note that the time delay and spatial separation between the EB and the spicule visibility are $\sim50$~s and $\sim1.0$~Mm, respectively. 
        The spatial separation lies within the interval reported in \citetalias{2025A&A...697A.180S}, while the delay in spicule visibility after the end of the EB is almost twice as long as the window we had set for our previous work.
        This means that we could potentially have found many more connections between EBs and spicules in the observational paper if we had chosen a more tolerant threshold for the time delay of visible spicules after the end of EBs near their footpoints.

    \subsection{Ellerman bombs as a tracer of shocks and reconnection}
    \label{disc: EB_tracer}

        Ellerman bombs are recognised as tracers of enhanced and localised heating in the lower solar atmosphere. 
        The energy required for such heating is especially significant in the deepest parts of the atmosphere, which is much denser than at higher altitudes. 
        As discussed in Sect.~\ref{meth: tracking_physics}, strong gradients are necessary to drive enhanced energy dissipation, resulting in localised heating. 
        This points to the action of shocks and/or current sheets.

        Ellerman bombs are now widely reported to be generated by magnetic reconnection, supported by both observations and simulations \citep{2002ApJ...575..506G, 2004ApJ...614.1099P, 2007A&A...473..279P, 2006ApJ...643.1325F, 2008ApJ...684..736W, 2017A&A...601A.122D, 2017ApJ...839...22H}. 
        Here, we present a more complex scenario in which both shocks and current sheets associated with a reconnection event can generate localised heating. 
        The EB emissions in our model trace this combined process.

        However, compression heating at the origin of such shock-induced EBs may be particularly prevalent in 2D simulations and may reflect a numerical bias arising from the simulation's numerical setup. 
        A reduction in the number of dimensions relative to 3D can artificially enhance the shock-heating mechanism \citep[see, e.g.][]{2005ApJ...631L.155U, 2012SoPh..276...75K}. 
        Therefore, the possibility of EBs induced solely by shock compression, without a current sheet, should be further investigated to determine whether this mechanism operates in the real solar atmosphere. 
        Progress will rely on more realistic numerical setups in 3D.

    \subsection{Radiative treatment to synthesise $\Ha$}
    \label{disc: limits}
    
        The 1.5D approximation is less accurate in the line core compared to the line wings of strong lines emitted in inhomogeneous environments \citep[such as, e.g. $\Ha$,][]{2012ApJ...749..136L}.
        However, EBs are observed in the line wings of $\Ha$.
        We therefore argue that the 1.5D approach is valid for tracking EBs at the spicule footpoint, which is the main objective of the radiative transfer in this work.

        To reduce computational cost when computing synthetic spectra, we used the standard RH 1.5D setup. 
        This setup calculates the hydrogen populations from the temperature and density profiles of the input atmosphere, assuming statistical equilibrium (SE), rather than using the detailed NEI populations from the Bifrost simulation. 
        This assumption affects the emergent intensity of the synthetic spectra \citep{2012ApJ...749..136L}. 
        However, \citet{2012ApJ...749..136L} conclude that SE can be assumed to calculate hydrogen spectra as long as the temperature and electron density in the model atmosphere are computed in a simulation that includes NEI for hydrogen. 
        Although we do not expect it to change our conclusions about the occurrence of EBs, direct comparisons with observations should be approached with care for the weakest profiles.

        For the strongest EB profiles at the spicule footpoint, the brightening goes well beyond the EB detection threshold of 145\% set by \citet{2013ApJ...774...32V} over multiple frames and columns.
        The EB profile with the strongest emission peaks within the chosen event (second panel of Fig.~\ref{fig: physics}) exceeds the average intensity by 264\%. 
        This immense brightening may have been slightly smaller if RH 1.5D used the NEI populations.

\section{Conclusions}
\label{conclusion}

    Our analysis has shown that the triggering of EBs and spicules can result from magnetic reconnection due to small-scale dynamics. 
    We showed in detail how one such magnetic reconnection event triggered both an EB and a spicule, causally connecting them. 
    Although we showed in \citetalias{2025A&A...697A.180S} that not all spicules have EBs at their footpoints, our results verify that the observed EB-spicule connections may be genuine and causal. 

    Our results provide strong arguments that we should not always expect to see flux emergence or cancellation in photospheric magnetic field maps at the sites of forming EBs. 
    Flux emergence and/or flux cancellation are the well-established processes forming the current sheets that trigger EBs. 
    Here, we present a new scenario for triggering EBs through small-scale dynamics: 
    An upward-propagating shock drags magnetic field lines towards an oppositely directed field. 
    It compresses a preexisting current layer into a reconnecting current sheet, which triggers the EB. 
    
    The same line of arguments applies to spicules, where we do not need flux emergence near the footpoints of forming spicules to trigger magnetic reconnection that drives spicules.
    We have shown that the shock's propagation amplifies magnetic tension along the current sheet, which is released when the advected magnetic field reconnects with the oppositely directed ambient field. 
    The reconnecting current sheet accelerates dense material from below through the reconnection jet, launching the long oblique spicule.

    In this work, we have used an rMHD simulation with self-consistent spicule generation to physically connect an EB to the formation of a type~II spicule. 
    We have also causally connected magnetic reconnection to type~II spicule formation. 
    Finally, we have shown that the small-scale dynamics of a shock can compress a thick current layer into a thin current sheet, triggering magnetic reconnection and leading to both an EB and a spicule.

\begin{acknowledgements}

This research is supported by the Research Council of Norway, project number 325491,
through its Centres of Excellence scheme, project number 262622,
and the European Research Council through the Synergy Grant number 810218 (`The Whole Sun', ERC-2018-SyG). 
The work of GA was supported by the Action Th\'ematique Soleil-Terre (ATST) of CNRS/INSU PN Astro, also funded by CNES, CEA, and ONERA. 
JMS acknowledges support by NASA contract NNG09FA40C (IRIS) and 80GSFC21C0011 (MUSE) and NSF grants AGS2532363 and AGS2532187. Bifrost simulations have been run on clusters, the Pleiades cluster, through the computing projects s1061, s2601 from the High-End Computing (HEC) division of NASA.
\end{acknowledgements}

\bibliographystyle{aa}
\bibliography{bibtex/sources}

\begin{appendix}

    \onecolumn

    \section{Bifrost simulation}

        \subsection{Simulation configuration}
        \label{subapp: sim_conf}
        
            The simulation we used in this work is presented in \citetalias{2020ApJ...889...95M}, and is based on the simulation presented by \citetalias{2017Sci...356.1269M}. 
            Both simulations include two terms of ion-neutral interactions in the induction equation: the Hall term and ambipolar diffusion, the latter of which is suggested as a key process for triggering type~II spicules. 
            \citetalias{2020ApJ...889...95M}'s simulation also includes NEI effects for hydrogen and helium. 
            The authors report that part of the energy typically used to ionise the plasma in local thermodynamic equilibrium is converted into thermal heating in NEI, leading to higher temperatures.

            The simulation includes a physics package regarding the different environments in the simulated solar atmosphere. 
            The convection zone, photosphere and lower chromosphere include full radiative transfer computations with a coherent scattering contribution in the gas emission \citep{2000ApJ...536..465S, 2010A&A...517A..49H}. 
            Radiative losses and gains for the upper chromosphere and the transition region follow \citet{2012A&A...539A..39C}. 
            The corona includes an optically thin radiative transfer and thermal conduction along the magnetic field \citep{2011A&A...531A.154G}. 
            An ad hoc heating term is also included to ensure that the temperature does not fall much lower than the equation of state's effective range ($\sim1800$~K).
    
            The simulation spans vertically from 2.5 Mm below the photosphere (upper convection zone) up to 40 Mm above the photosphere (corona), and horizontally over 96 Mm. 
            The vertical resolution is nonuniform, ranging from $\sim12$ km to $\sim70$ km; it is smallest in the photosphere, chromosphere and transition region, and increases in the convection zone and corona. 
            The horizontal resolution is uniform at $\sim14$ km.
            
        \subsection{Classification as type II spicule}
        \label{subapp: type II}
        
            As type~II spicules in observations of active regions have Doppler excursions similar to type I spicules, the fading of the spicule in the $\Ha$ line is the key characteristic to classify the selected spicule as a type~II. 
            This is demonstrated in the $\lambda t$~diagram in Fig.~\ref{fig: lambda_t}. 
            To further establish that the fading in $\Ha$ is not only visible for the selected column B, we made a series of $\lambda x$ diagrams.
    
\begin{figure}
        \resizebox{\hsize}{!}{\includegraphics{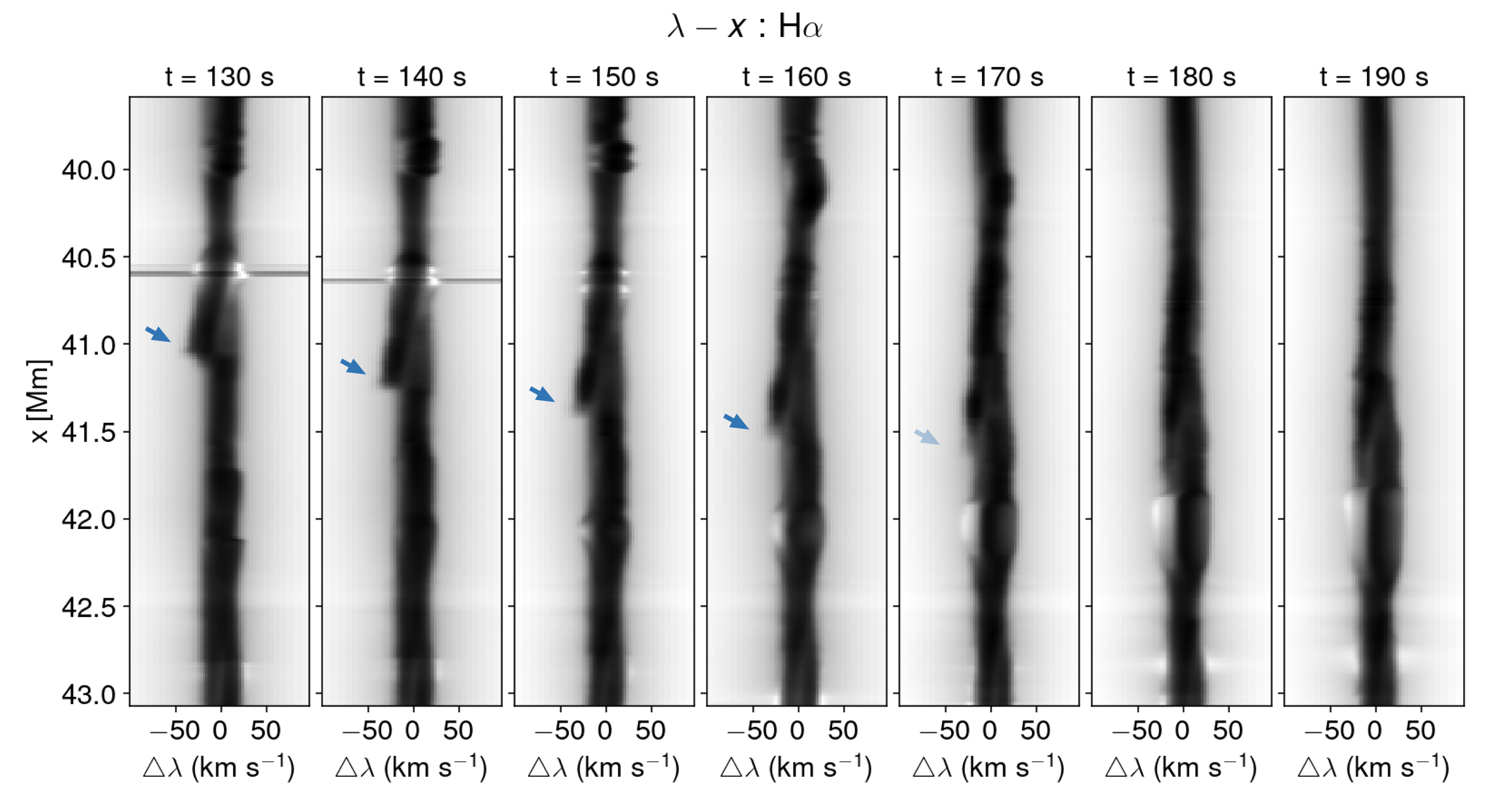}}
        \caption{
            Fading of the spicule in $\Ha$ in a series of $\lambda x$~diagrams. 
            The $\Ha$ blue-shifted excursion of the spicule is indicated by the blue arrows, which are fading out together with the spicule.
            The spatial range along the $y$-axis is the same as in Fig.~\ref{fig: spicule}.
        }
        \label{fig: lambda_x}
        \end{figure}

            The series of $\lambda x$~diagrams is presented in Fig.~\ref{fig: lambda_x}, and follows the evolution of the spicule over seven time steps with a 10~s interval. 
            The $y$-axes of these diagrams cover the same spatial range in $x$ as Fig.~\ref{fig: spicule}, which captures the entire spicule during its lifetime. 
            The $\lambda x$~diagrams show that the spicule starts to fade out at about $t=160$~s, before it is no longer visible at 170~s. 
            The spicule is no longer visible in the blue wing of $\Ha$, yet it continues to rise, demonstrating that the spicule has faded and is no longer visible in the chromospheric line.
    
    \section{Contribution function analysis}
    \label{app: cont_func}

        The contribution function analysis is a 2-dimensional plot of the contribution to the emergent intensity of a spectral line, showing the dependence on wavelength along the $x$-axis and on height along the $y$-axis. 
        \citet{Carlsson_1997} arrive at the contribution function from the formal solution of the transfer equation, and it may be written as 
        \begin{equation*}
            C_{I}(\lambda)\ =\ \dfrac{\text{d}I_{\lambda}}{\text{d}z}\ =\ \dfrac{\chi_{\lambda}}{\tau_{\lambda}}\ \cdot\ S_{\lambda}\ \cdot\ \tau_{\lambda} e^{-\tau_{\lambda}},
        \end{equation*}
        where $\chi_{\lambda} / \tau_{\lambda}$ is the ratio of opacity to optical depth, $S_{\lambda}$ is the total source function (continuum and line), $\tau_{\lambda} e^{-\tau_{\lambda}}$ is a factor highlighting where $\tau \approx 1$.

        The first factor, $\chi_{\lambda} / \tau_{\lambda}$, becomes significant in places where there are many emitting particles (large $\chi_{\lambda}$) at small optical depths (small $\tau_{\lambda}$). 
        The total source function, $S_{\lambda}$, is important where the density is high enough to couple the source function to the Planck function. 
        Together with the last factor, $\tau_{\lambda} e^{-\tau_{\lambda}}$, we can identify the main contributors to the emergent intensity as a function of wavelength and height. 
        The four-panel contribution function analysis for the strongest EB of our chosen event is presented in Fig.~\ref{fig: cont_func}.

    \section{Flux emergence}
    \label{app: flux_emer}

        To look for emergence of new flux into the atmosphere related to our event, we examine the horizontal magnetic field strength below and above the photosphere, centred around the event. 
        There is a flux emergence episode at $x = 39.5$~Mm, where a flux rope emerges through the photosphere at the onset of the EB. 
        The emerging magnetic flux is located at the boundary between the convection zone and the photosphere when the EB starts, i.e. it is at least 0.6~Mm separated from the reconnection site. 
        This already indicates that the emergence of new flux does not drive the reconnection.

        This is further supported by the dynamics of the horizontal magnetic field in the area of the propagating magnetoacoustic shock (this is especially evident in the corresponding animation). 
        From $t = -40$~s, i.e. when the shock is first captured by the criterion of Eq.~\ref{eq: alpha_crit}, to 20~s, i.e. when the current sheet is stationary and reconnecting, it is evident that the magnetic field strength is following the dynamics of the shock. 
        In this timeframe, the magnetic field, as illustrated by the horizontal magnetic field strength, is falling just above the reconnection site, following the motion of the downflowing plasma towards the magnetic field advected by the shock. 
        This suggests that the current layer may be compressed even faster due to the opposing motion of the oppositely directed magnetic fields, aiding the build-up of magnetic tension and subsequent magnetic reconnection.

        We conclude that, in fact, magnetic flux is emerging below our chain of events that trigger the EB and the spicule, but that these are two separate events in the simulation.
    
        \begin{figure}
        \resizebox{\hsize}{!}{\includegraphics{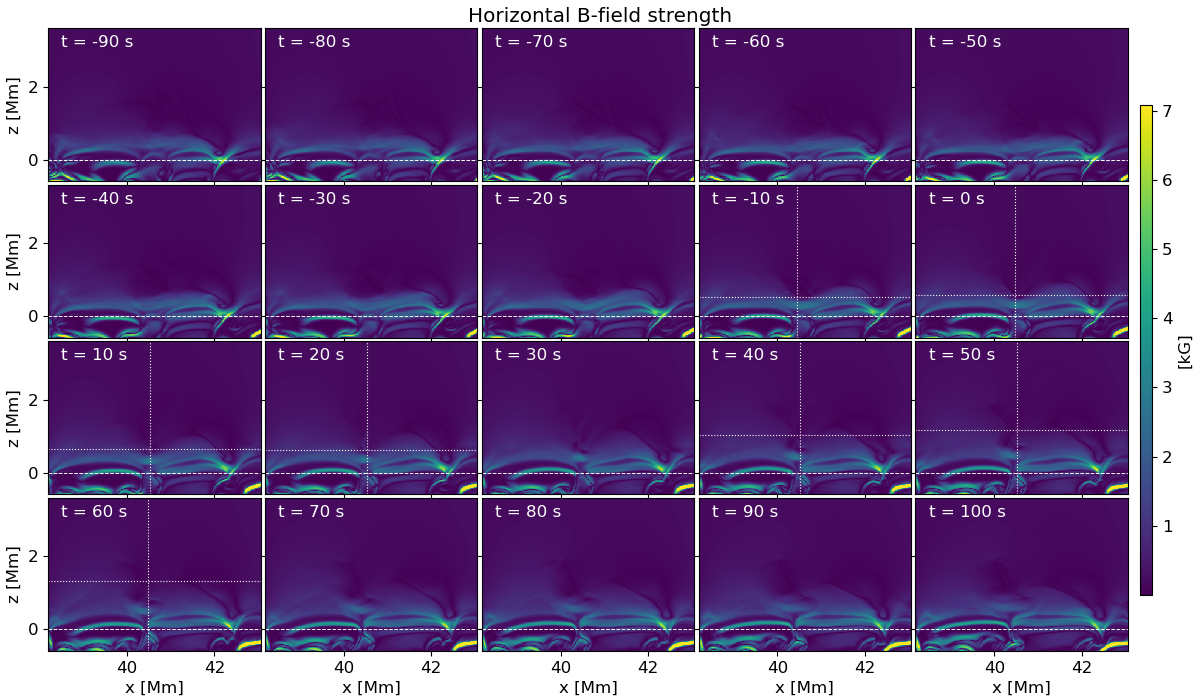}}
        \caption{
            Temporal evolution of the horizontal magnetic field strength centred around the current sheet before, during, and after the EB. 
            The white dashed line marks $z = 0$~Mm, i.e. the photosphere. 
            The white dotted lines mark the location of the EB at their intersection. 
            An animation of this figure is available at \url{http://tsih3.uio.no/lapalma/subl/recon_eb_spic/flux_emergence.mp4}.
        }
        \label{fig: bh_photo}
        \end{figure}

    \twocolumn
    
\end{appendix}

\end{document}